\documentclass[manuscript]{aastex}

\usepackage{epsfig}
\usepackage{times}
\usepackage{natbib}
\usepackage{color}
\usepackage{rotating}
\usepackage{graphicx}

\newcommand{\gps}{\ensuremath{g_{\rm P1}}}
\newcommand{\rps}{\ensuremath{r_{\rm P1}}}
\newcommand{\ips}{\ensuremath{i_{\rm P1}}}
\newcommand{\zps}{\ensuremath{z_{\rm P1}}}
\newcommand{\yps}{\ensuremath{y_{\rm P1}}}

\title{Properties of M31. IV: Candidate Luminous Blue Variables from PAndromeda}
\author{C.-H. Lee\altaffilmark{1,2}, S. Seitz\altaffilmark{1,2}, M. Kodric\altaffilmark{1,2}, A. Riffeser\altaffilmark{1,2}, J. Koppenhoefer\altaffilmark{2,1}, R. Bender\altaffilmark{1,2}, J. Snigula\altaffilmark{2,1}, U. Hopp\altaffilmark{1,2}, C. G\"ossl\altaffilmark{1,2}, L. Bianchi\altaffilmark{3}, P. A. Price\altaffilmark{4}, M. Fraser\altaffilmark{5}, W. Burgett\altaffilmark{6}, K. C. Chambers\altaffilmark{6}, P. W. Draper\altaffilmark{7}, H. Flewelling\altaffilmark{6}, N. Kaiser\altaffilmark{6}, R.-P. Kudritzki\altaffilmark{6}, E. A. Magnier\altaffilmark{6}}
\altaffiltext{1}{University Observatory Munich, Scheinerstrasse 1, 81679 Munich, Germany}
\altaffiltext{2}{Max Planck Institute for Extraterrestrial Physics, Giessenbachstrasse, 85748 Garching, Germany}
\altaffiltext{3}{Department of Physics and Astronomy, Johns Hopkins University, 3400 North Charles Street, Baltimore, MD 21218, USA}
\altaffiltext{4}{Department of Astrophysical Sciences, Princeton University, Princeton, NJ 08544, USA}
\altaffiltext{5}{Astrophysics Research Centre, School of Mathematics and Physics, Queen's University Belfast, Belfast BT7 1NN, UK}
\altaffiltext{6}{Institute for Astronomy, University of Hawaii at Manoa, Honolulu, HI 96822, USA}
\altaffiltext{7}{Department of Physics, Durham University, South Road, Durham DH1 3LE, UK}  

\begin{document}

\begin{abstract}
We perform a study on the optical and infrared photometric properties of known 
luminous blue variables (LBVs) in M31
using the sample of LBV candidates from the Local Group Galaxy Survey \citep{2007AJ....134.2474M}. 
We find that M31 LBV candidates show photometric variability ranging 
from 0.375 to 1.576 magnitudes in $\rps$ during a three year time-span observed
by the Pan-STARRS 1 Andromeda survey (PAndromeda). 
Their near-infrared colors also follow the distribution
of Galactic LBVs as shown by \cite{2013A&A...558A..17O}.
We use these features as selection criteria to search for unknown LBV candidates in M31.
We thus devise a method to search for candidate LBVs using both optical 
color from the Local Group Galaxy Survey and infrared color from 
Two Micron All Sky Survey, as well as photometric variations observed 
by PAndromeda.
We find four sources exhibiting common properties of 
known LBVs. These sources also exhibit UV emission as seen 
from \textit{GALEX}, which is one of 
the previously adopted method to search for LBV candidates. 
The locations of the LBVs are well aligned with 
M31 spiral arms as seen in the UV light, suggesting they are evolved stars at 
young age given their high-mass nature.  
We compare these candidates with the latest Geneva evolutionary 
tracks, which show that our new M31 LBV candidates 
are massive evolved stars with an age of 10 to 100 million years.

\end{abstract}


\section{Introduction}
Luminous blue variables are hot massive stars which undergo sporadic eruptions on 
timescales of years and decades \citep{1994PASP..106.1025H}. 
The prototype is S Doradus, as well as Hubble-Sandage variables 
in M31 and M33 \citep{1953ApJ...118..353H}, which shows eruptions of 1-2 magnitude 
level in a time-span of several decades. Other examples are $\eta$ Carina and 
P Cygni, which show giant eruptions ($>$ 2 mag) at a frequency of several centuries.
\cite{1984IAUS..105..233C} is the first to coin the name Luminous Blue Variables
for this type of stars, and separates them from other type of bright blue stars like Wolf-Rayet stars.

LBVs play an important role at the very late stage of massive star evolution. They are considered as a 
transition phase where O stars evolve toward Wolf-Rayet Stars \citep{2011A&A...525L..11M}. LBVs were 
originally regarded only as supernova impostors because they often show giant eruptions mimicking the explosion of 
supernovae, but the central star remains after the ejecta have been expelled. 
However, a link between LBVs and supernova progenitors was suggested 
by \cite{2006A&A...460L...5K} when interpreting
the radio lightcurves of supernovae. The radio emission seen after the supernova explosion is induced 
by the interaction between supernova ejecta and the progenitor's circumstellar medium, thus radio lightcurves 
bear information on the 
mass-loss history of the progenitor. \cite{2006A&A...460L...5K} suggested that radio 
lightcurves of SNe indicate high mass-loss histories of the progenitors which matches well with LBVs. 

Pre-eruption images of several SNe also suggest LBVs as their progenitor. 
For example, the progenitor of SN 1987A was recognized as a blue supergiant \citep{1987ApJ...321L..41W} 
and \cite{2007AJ....133.1034S} suggested that
it could be classified as a low-luminosity LBV. \cite{2009Natur.458..865G} identified the progenitor 
of SN 2005gl using \textit{HST} and attributed it to be a LBV. Recently a previously known LBV - SN 2009ip - 
has undergone its third eruption and has been linked to a true supernova  \citep{2013MNRAS.430.1801M}.
The nature of the recent eruption of SN 2009ip is under debate; subsequent follow-up has been carried out to verify or reject 
it as a core-collapse SN \citep{2013MNRAS.433.1312F}. Yet there are only a few known LBVs, either in our Galaxy or in M31 and M33. 
Thus, increasing the number of known LBVs is 
essential toward understanding their nature and evolution.

In addition to the pioneering decades-long photometric monitoring campaign conducted by Hubble and Sandage \citep{1953ApJ...118..353H}, 
there are several methods to uncover LBVs. For example, LBVs are strong UV and H-alpha emitters 
\citep[see][ and reference therein]{2007AJ....134.2474M} and can be revealed e.g. with  
observations of the \textit{GALEX} satellite 
or H-alpha surveys. \cite{2007AJ....134.2474M} conducted
a H-alpha survey of M31 and M33 and followed-up a selected sample of strong H-alpha emitters 
spectroscopically. By comparing 
the spectra of their H-alpha emitter sample with known LBVs, they were able to identify candidate LBVs, 
which saved a substantial amount of
time required to uncover LBVs photometrically. Because they have uncovered more than 2,500 H-alpha emitting stellar
objects, they can only follow-up dozens of them, yet there are much more to be explored. \cite{2013ApJ...773...46H} are currently 
exploring other H-alpha emitting sources in this list, in combination with infrared photometry including 
2MASS, \textit{Spitzer} 
and \textit{WISE} to search for luminous and variable stars. 
Since LBVs undergo several eruptions and exhibit high mass-loss rates, they accumulate vast amounts 
of material in their circumstellar environment which could be detectable in the infrared \citep[e.g.][]{2012MNRAS.421.3325G}.
\cite{2013ApJ...767...52K} have made use of \textit{Spitzer} IRAC photometry and searched for $\eta$ Carina analogs in nearby galaxies 
including M33 (but not M31). They estimate that 6$\pm$6 of their candidates are true $\eta$ Carina-like sources.

Here we outlined a novel approach utilizing mid-term 
photometric variation, as well as optical and infrared color to search 
for LBVs using the 
LGGS optical and 2MASS infrared photometry,
with the combination of the photometric variability from PAndromeda 
monitoring campaign. 
Our paper is organized as follows: in \textsection~2 we describe the optical and infrared data we use. 
In \textsection~3 we outline our method. A discussion of our candidates is presented in \textsection~4, 
followed by an outlook in \textsection~5. 
\\

\section{Data sample}

\subsection{Optical data}

\begin{figure*}[!h]
  \centering
  \includegraphics[scale=1.0]{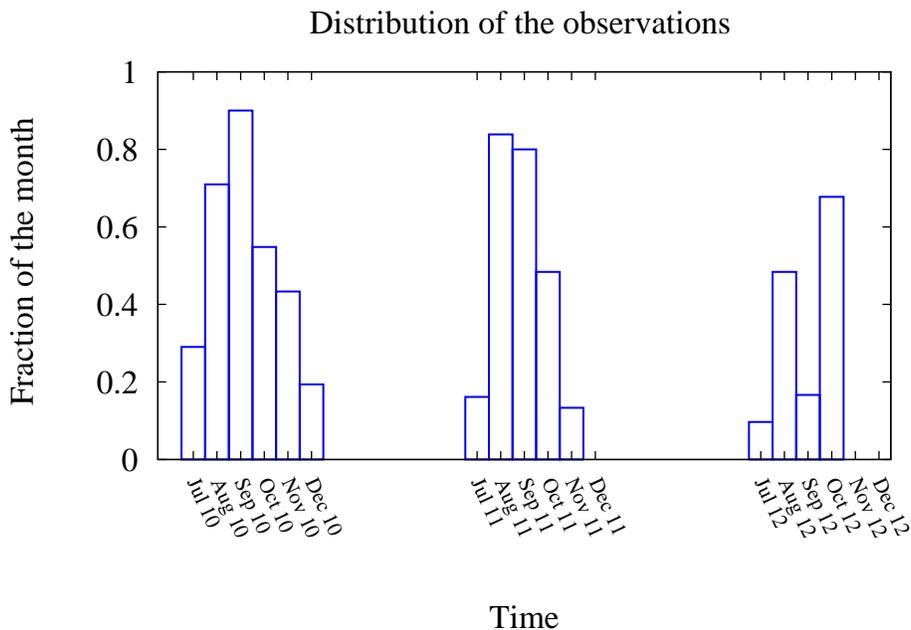}
  \caption{The distribution of the observations of PS1 towards M31. We plot the monthly fraction of nights in the $\rps$-filter during the 2010, 2011 and 2012 seasons. In general, the observations cover most of the time in the second half of each year.}
   \label{fig.cadence}
\end{figure*}

The time-series photometric data employed to search for variability are from the PAndromeda project.
PAndromeda monitored the Andromeda galaxy with the 1.8m PS1 telescope with a $\sim$ 7 deg$^2$ 
field-of-view \citep[see][ for a detailed description of the PS1 system, optical design, and the imager]{2010SPIE.7733E..12K,2004SPIE.5489..667H,2009amos.confE..40T}.
Observations have been taken in the $\rps$ and $\ips$ filters on daily basis during July to 
December between 2010 and 2012
in order to search for microlensing events and variables. The distribution of the observations in the $\rps$-filter is shown in Fig. \ref{fig.cadence}. 
Several exposures in $\gps$, $\zps$, and $\yps$ 
are also taken as complementary information for studies of the stellar content of M31.

The data reduction is based on the MDia tool \citep{2013ExA....35..329K} and is explained in \cite{2012AJ....143...89L} in detail. 

We outline our data reduction 
steps as follows. 
The raw data are detrended by the image processing pipeline \citep[IPP,][]{2006amos.confE..50M} and 
warped to a sky-based image plane (so-called skycells). 

The images at the skycell stage 
are further analyzed by our sophisticated imaging subtraction pipeline \textit{mupipe} \citep{2002A&A...381.1095G}
based on the idea of image differencing analysis advocated by \cite{1998ApJ...503..325A}. This includes the creation of deep
reference images from best seeing data, stacking of observations within one visit to have better signal
to noise ratio (hereafter ``visit stacks''), subtraction of visit stacks from the reference images
to search for variabilities, and creating lightcurves from the subtracted images.  

We have shown in \cite{2013AJ....145..106K}  how to obtain lightcurves for resolved sources from the PAndromeda data. 
The major difference of the data used in this work is that our present data-set
contains three years of PAndromeda data, instead of one year and a few days from the 
second year data used in \cite{2013AJ....145..106K}.
The sky tessellation is also different, in order to have the central region of M31 in the center of a skycell 
(skycell 045), instead of at the corner of adjacent skycells (skycell number 065, 066, 077, and 078) as in 
\cite{2013AJ....145..106K};
the skycells are larger and overlap in the new tessellation. The new tessellation is drawn in Fig. 1 of \cite{2013ApJ...777...35L}. 
We have extended the analysis to 47 skycells, twice as many as the number of skycells used in 
\cite{2013AJ....145..106K}. 
The skycells we used are 012-017, 022-028, 032-038, 042-048, 052-058, 062-068, 072-077, which cover an area of 7 deg$^2$ 
of M31. 
The search of variability is conducted in both $\rps$ and $\ips$, where we start from the resolved sources in 
the $\rps$ reference images, and check for variability in both $\rps$ and $\ips$ filters. 

In addition, we also use the deep photometric catalog from the Local Group Galaxy Survey 
\citep[LGGS, ][]{2006AJ....131.2478M}. The LGGS 
utilized the 4-meter KPNO telescope to observe the M31 galaxy. Their Mosaic CCD camera has a resolution 
of 0''.261 / pixel at the center which decreases to 0''.245 / pixel towards the corner. The field-of-view of the 
camera is roughly 36' $\times$ 36'. M31 was observed between 2000 and 2002 in Johnson \textit{UBVRI} 
filters with seeing values from 0''.8 to 1''.4. The observations cover ten fields, 
corresponding to 2.2 deg$^2$ along the major axis of M31 \citep[see Fig. 1 of][]{2006AJ....131.2478M}. 

The astrometric solution for each frame was derived by matching with the USNO-B1.0 catalog 
\citep{2003AJ....125..984M}. The \textit{PSF} photometry in each filter was obtained with the IRAF DAOPHOT 
routine, and calibrated against the Lowell 1.1m data with zero points and color terms. The final LGGS catalog 
contains 371,781 stars in M31, with 1 - 2 percent statistic error at 21 mag and 10 percent at 23 mag. When matching the LGGS catalog to the PAndromeda catalog, we found an median astrometric difference of 0.36 arcsec.

\subsection{Infrared data}

In this work we utilize the catalog and images delivered by the Two Micron All Sky Survey 
\citep[2MASS, ][]{2006AJ....131.1163S}. 2MASS employed two 1.3-meter telescopes located at Mt. Hopkins, 
Arizona and Cerro Tololo, Chile to survey the full sky in three near infrared passbands J(1.25$\mu$m), 
H(1.65$\mu$m), and K$_s$(2.16$\mu$m) simultaneously. The pixel scale of the 2MASS CCDs is 2 arcsec / pixel. 
2MASS observed every patch of the sky with six times 1.3 s integration time. The raw images were dark subtracted, 
flat-fielded, sky subtracted, resampled to a 1 arcsec /pixel coordinate grid in a flux-conserving manner, and 
coadded to generate an Atlas Image.

The source detection was performed on the Atlas Images with PSF profile-fitting, yielding a sensitivity of 15.8, 15.1, and 
14.3 mag at 10 $\sigma$ level in J, H, and K$_s$ band, respectively. The astrometry was calibrated against the Tyco-2 
Reference Catalog \citep{2000A&A...357..367H} and yielded an order of 100 mas accuracy for bright sources.

We queried the 2MASS point-source catalog for a 3 $\times$ 3 deg$^2$ area centered at M31 via the 
NASA/IPAC Infrared Science Archive, and retrieved 43,723 sources in this region. In addition, we also
retrieved postage-stamp images from IRSA to examine the sources.

\section{Selection method}
We designed our selection algorithm based on the optical and infrared properties of known 
LBVs. 

\begin{figure*}[!h]
  \centering
  \includegraphics[scale=0.6]{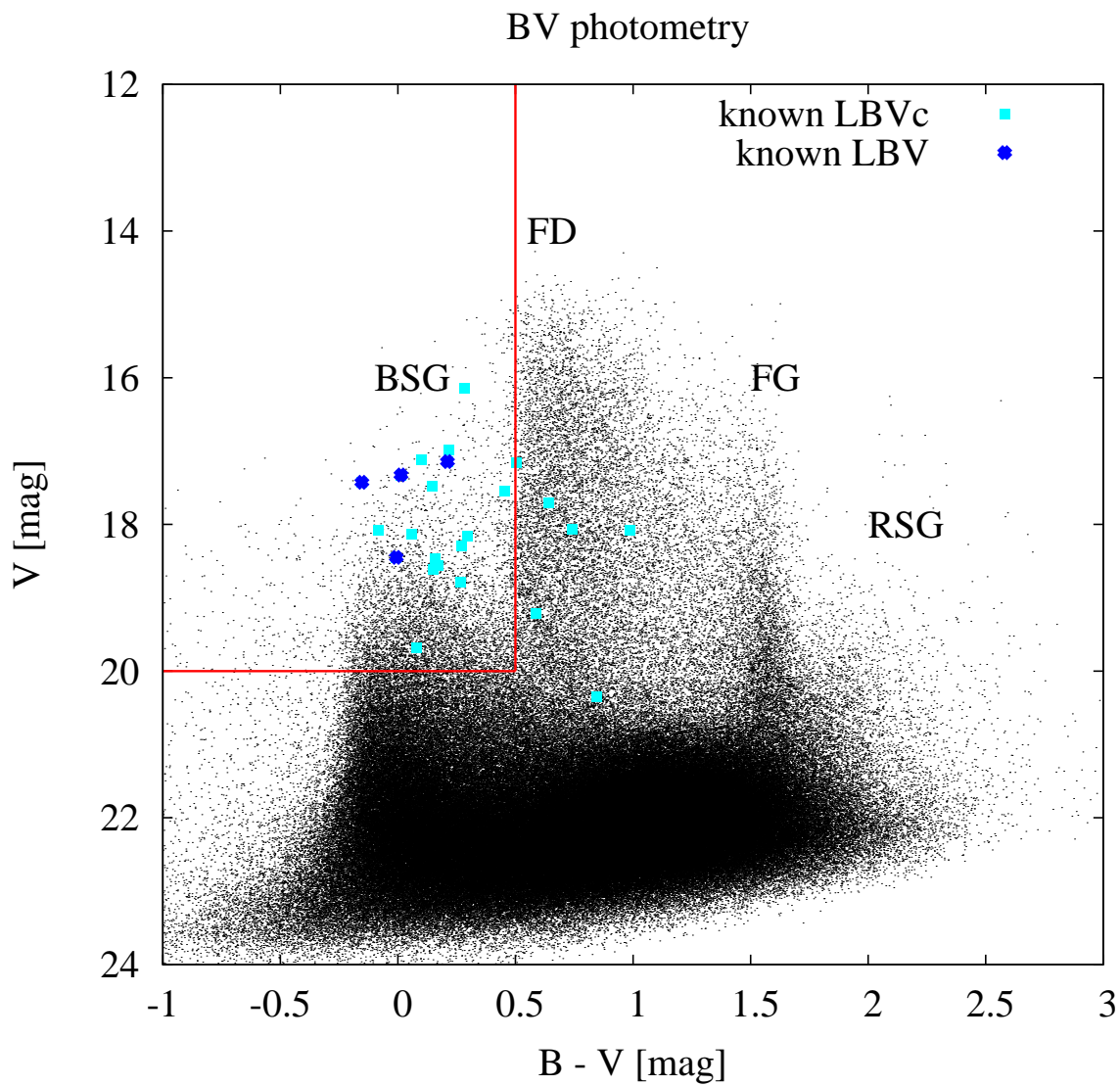}
  \caption{M31 color-magnitude diagram using the \textit{B} and \textit{V} photometry from \cite{2006AJ....131.2478M}. All sources in their catalog are shown with black dots. The peaks at \textit{B - V} $\sim$ 0.0, 0.6, and 1.5 are from blue supergiants (BSG), foreground dwarfs (FD), and foreground giants (FG), respectively. We thus make a cut at \textit{B - V} $<$ 0.5 and V brighter than 20 mag (marked with red lines) to avoid contamination from foreground stars and RSGs. The known LBVs in the LGGS sample are plotted with blue points, the LBV candidates listed in the LGGS sample are plotted with cyan points.} 
  \label{fig.criteria_VBV}
\end{figure*}

The first criterion is the optical color. As can be seen in Fig. \ref{fig.criteria_VBV}, 
the known LBVs and LBV candidates presented by \cite{2007AJ....134.2474M} 
are rather blue in the B-V v.s. V color-magnitude diagram (CMD). We thus set the criterion 
that B-V $<$ 0.5 mag and V brighter than 20 mag to select for optically blue and luminous 
objects. This enables us to filter out possible contaminations from 
foreground stars and unresolved background galaxies.

The next criterion utilizes the optical variability from PAndromeda data. Since OB stars also show aperiodic variation
at $<$ 0.1 magnitude level - the so-called $\alpha$ Cyg variables \citep{1998A&AS..128..117V} - and since \cite{2012A&A...541A.146C} 
have accounted for
variation $\leq$ 0.4 magnitude in their M33 LBV candidates as by similar mechanism, we require lightcurve variations $>$ 0.4 mag 
from $\rps$ to secure mid-term LBV variability. Loosing this criterion might help to gain more candidates. We will come to this
point in section \ref{sec.det_eff}. Sometimes there are a few data points with huge error-bars which deviate from other observations. Such outlier
measurements could render a lightcurve with $\Delta$ mag $>$ 0.4 and leads to false detection. To reduce the number of false detection, we 
thus require that in a single lightcurve, at least 25 data points vary at 10-$\sigma$ level with respect to the mean value from all the 
measurements.   

\begin{figure*}[!h]
  \centering
  \includegraphics[scale=1.0]{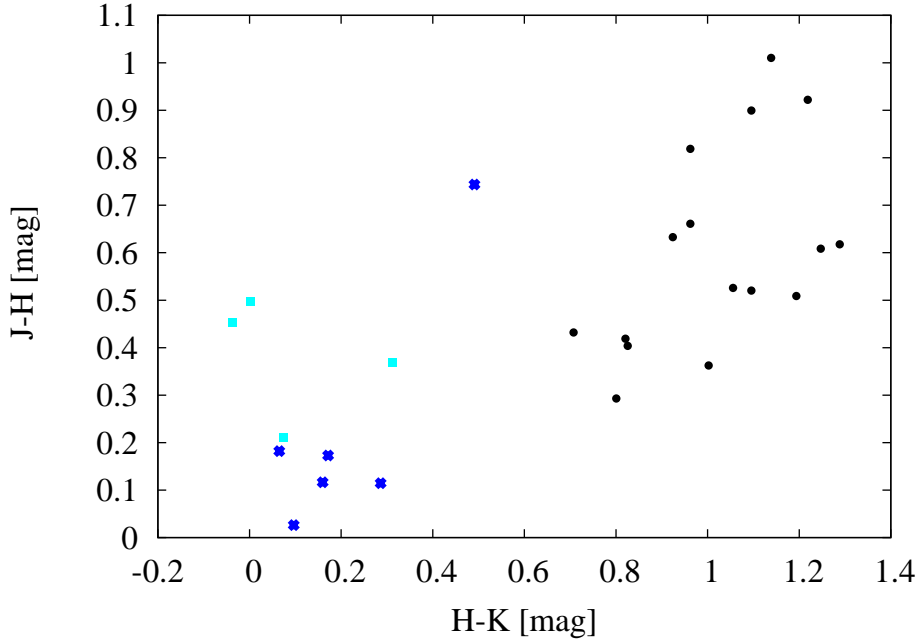}
  \caption{Infrared color-color diagram of LBVs and B[e]SGs using the 
2MASS photometry from \cite{2006AJ....131.1163S}. 
Galactic LBVs in the \cite{2013A&A...558A..17O} sample are shown in blue color. 
Known M31 LBVs are shwon in cyan color. The Galactic B[e]SG sample from 
\cite{2013A&A...558A..17O} 
are shown in black, which distribute in a region at H-K $>$ 0.5 and are 
distinct from the LBVs. }
  \label{fig.jhk}
\end{figure*}

The last criterion is set by the infrared color. \cite{2014ApJ...780L..10K} have indicated that B[e] super
giants (B[e]SGs) also show variabilities similar to LBVs, and pointed out that LBVs and B[e]SGs can be distinguished
from the infrared colors as shown in a recent study by \cite{2013A&A...558A..17O} using samples from the Milky Way and 
the Magallanic Clouds. We plot the sample of \cite{2013A&A...558A..17O} in Fig. \ref{fig.jhk}, as well as the four known 
LBVs in M31. The M31 LBVs follow the distribution of Galactic LBVs; most of them have H-K $<$ 0.5. We thus 
set a upper H-K limit of 0.5 to select candidate LBVs.

There are seven sources pass the aforementioned criteria. We inspect their 
lightcurves individually and select four as candidate LBVs. 
For the remaining three sources, they are all known LBVs (AE And, M31 Var 15, and M31 Var A-1). 
The only known LBV that does not pass our selection criteria is AF And, due to its variability smaller than the 0.4 mag criterion
(0.375) during the time-span of PS1.

\begin{figure*}
  \centering
  \includegraphics[scale=0.8]{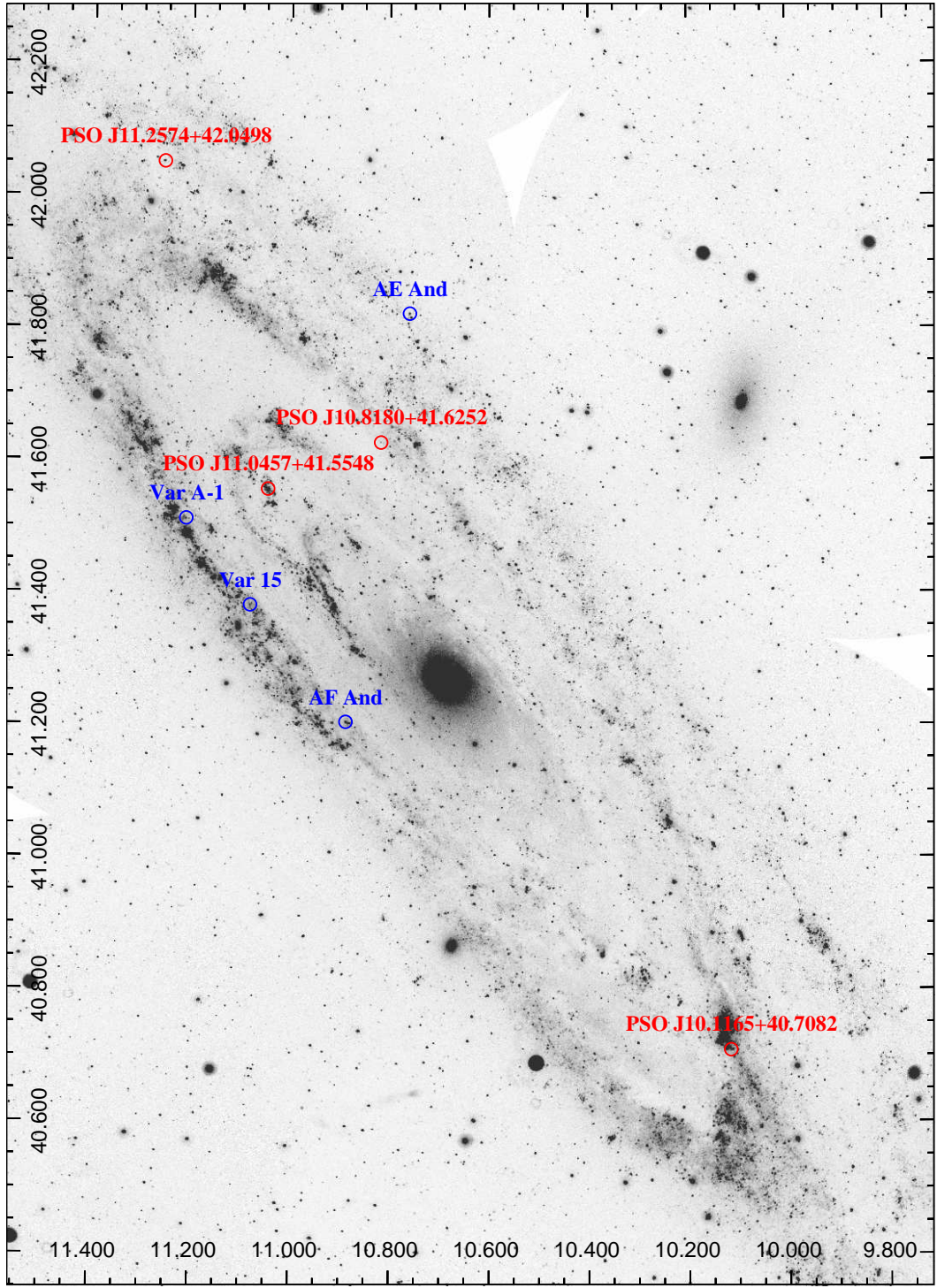}
  \caption{Spatial distribution of our LBV candidates (red circles) and the LBVs listed in \cite{2007AJ....134.2474M} (blue circles). 
The underlying image is the \textit{GALEX} near UV map by \cite{2007ApJS..173..185G}.}
   \label{fig.spat}
\end{figure*}

\begin{figure*}
  \centering
  \includegraphics*[scale=0.6]{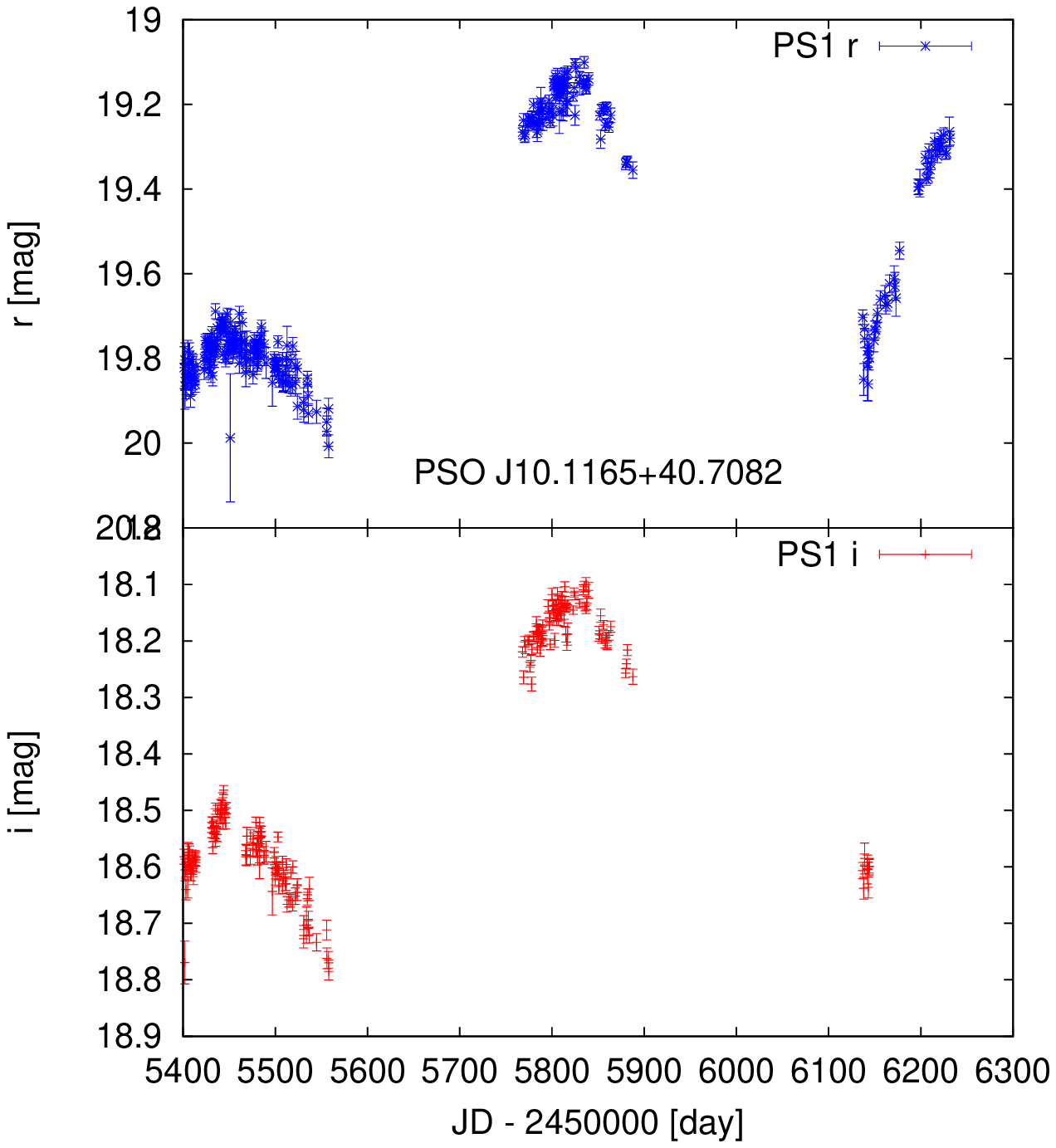}
  \includegraphics*[scale=0.6]{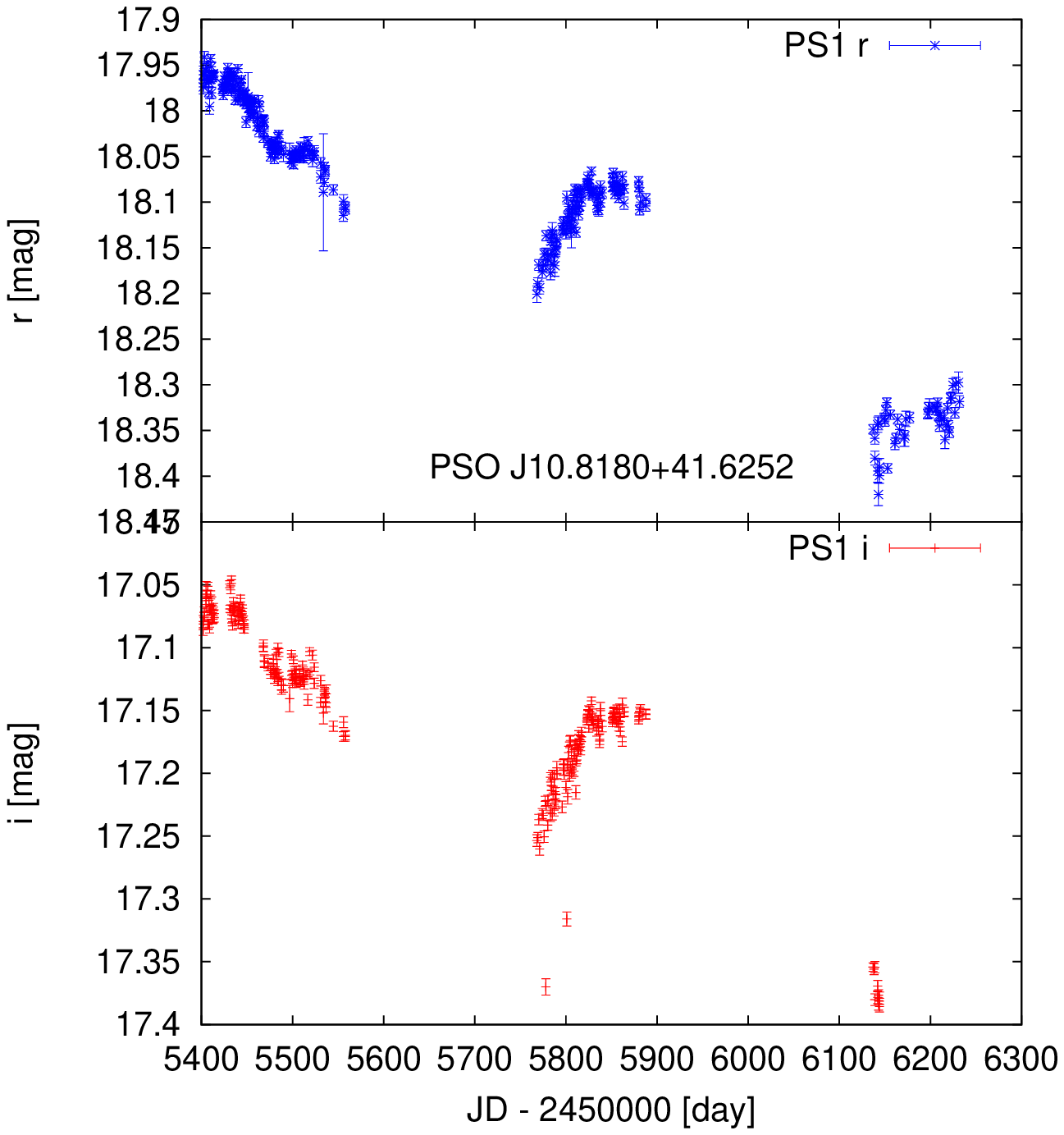}
  \includegraphics*[scale=0.6]{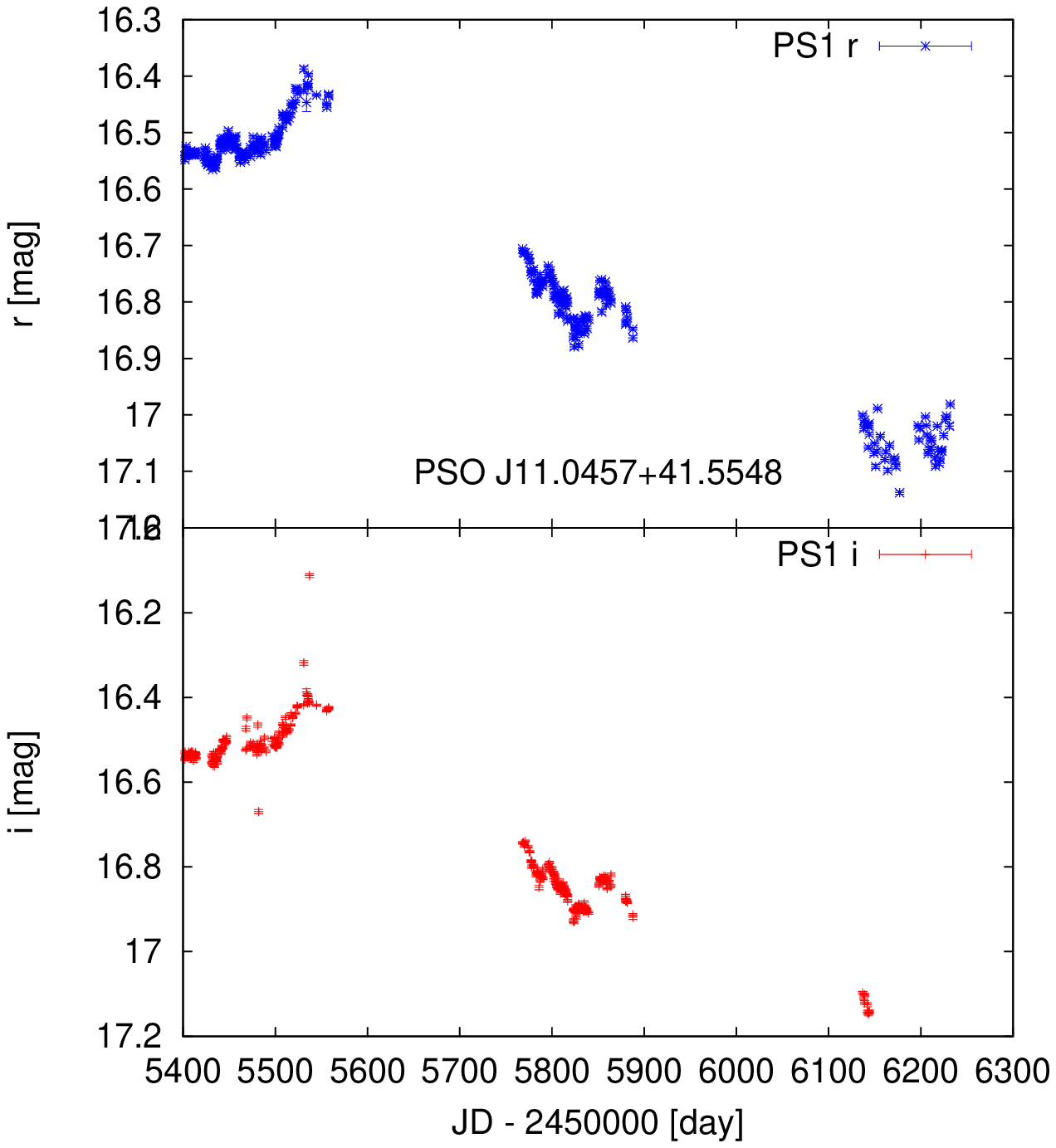}
  \includegraphics*[scale=0.6]{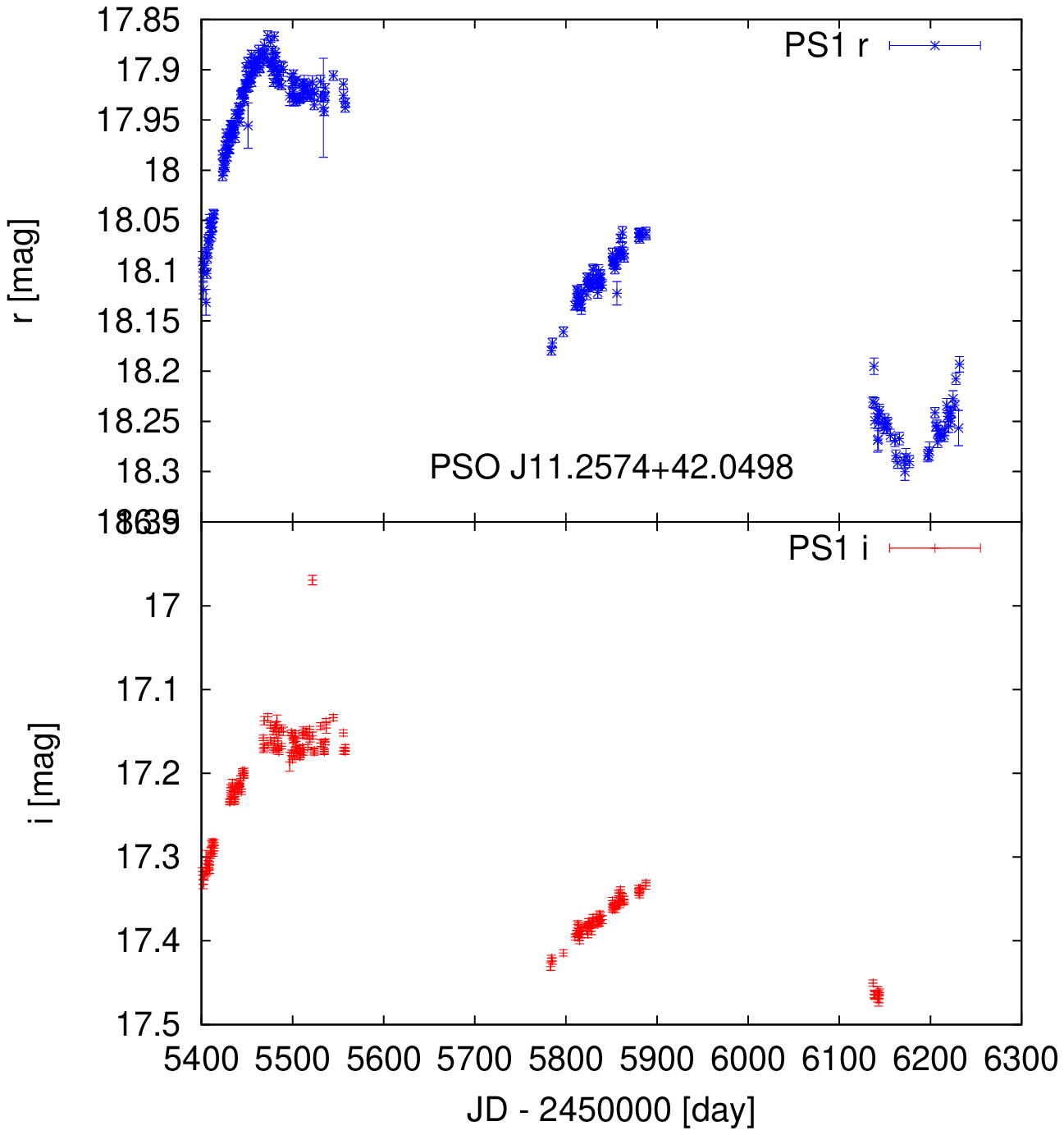}
  \caption{Lightcurves of our LBV candidates. From left to right, top to 
bottom are: PSOJ10.1165+40.7082, PSOJ10.8180+41.6252, PSOJ11.0457+41.5548, 
and PSOJ11.2574+42.0498. The lightcurves are obtained 
from the PS1 PAndromeda survey. The blue and red points 
indicate the $\rps$ and $\ips$ observations, respectively.}
   \label{fig.opt}
\end{figure*}

The number of stars that passed each criterion is listed in Table \ref{tab.crit}. 
There are in total four previously unknown sources which pass all our criteria. 
Their optical and infrared photometry, as well as their spatial distribution and lightcurves, 
are shown in Table \ref{tab.cphot}, Fig. \ref{fig.spat} and \ref{fig.opt}, respectively.

We discuss these sources in the following section. 

\begin{table}
\center
\begin{tabular}[t]{lr}
\hline 
Criterion  & Number of passed stars \\
\hline
Total LGGS sources & 371781 \\
B-V $<$ 0.5 and V $<$ 20 & 5928 \\
$\Delta$mag $>$ 0.4 & 9 \\
H-K $<$ 0.5 & 7 \\
Previously unknown & 4\\
\hline
\end{tabular}
\caption{Number of stars passed in each criterion.}
\label{tab.crit}
\end{table}

\begin{table*}
\centering
\begin{tabular}[t]{llllllll}
\hline 
Name  & RA          & Dec         & V      & B-V   & J  & H  & K$_s$ \\
      & (J2000)     & (J2000)     &        &       &    &    &   \\
\hline
PSO J10.1165+40.7082 & 00:40:28.00 & +40:42:29.1 & 19.513 & 0.027 & 15.617 & 14.729 & 14.319 \\ 
PSO J10.8180+41.6265 & 00:43:16.33 & +41:37:30.6 & 19.494 & 0.164 & 14.671 & 13.896 & 13.458 \\ 
PSO J11.0457+41.5548 & 00:44:11.01 & +41:33:17.6 & 17.300 & 0.042 & 16.483 & 15.085 & 14.884 \\ 
PSO J11.2574+42.0498 & 00:45:01.84 & +42:02:59.2 & 18.498 & 0.344 & 15.244 & 14.501 & 14.146 \\ 

\hline
\end{tabular}
\caption{Optical and infrared photometry of our LBV candidates. B and V photometry are taken from the LGGS; 
\textit{JHK$_s$} photometry are taken from the 2MASS catalog \citep{{2006AJ....131.1163S}}}
\label{tab.cphot}
\end{table*}
\begin{figure*}
  \centering
  \includegraphics[scale=0.9]{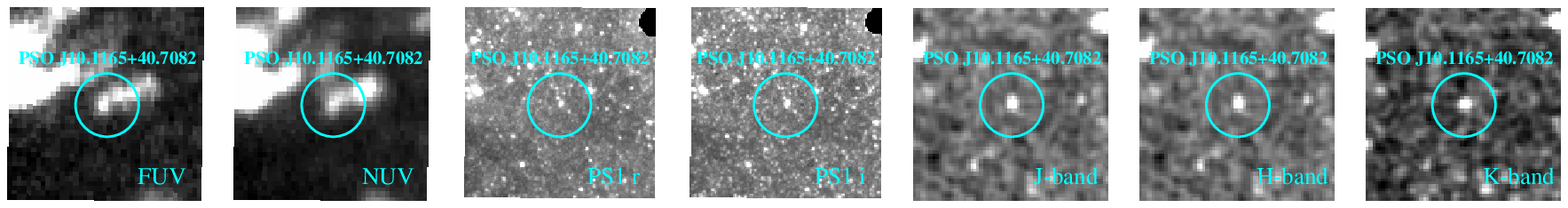}  
  \includegraphics[scale=0.9]{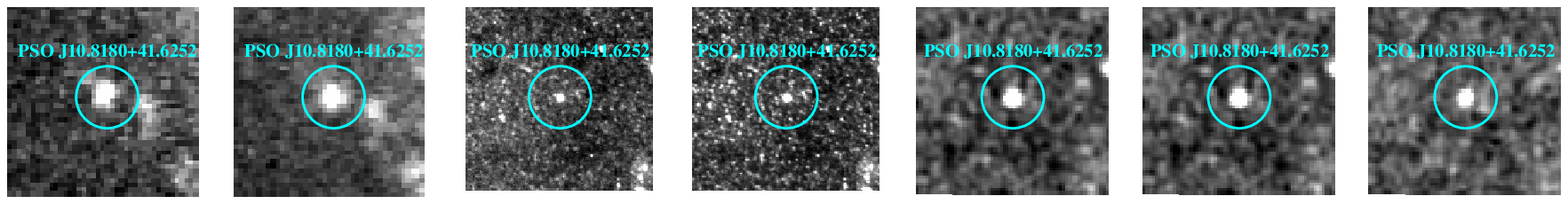}  
  \includegraphics[scale=0.9]{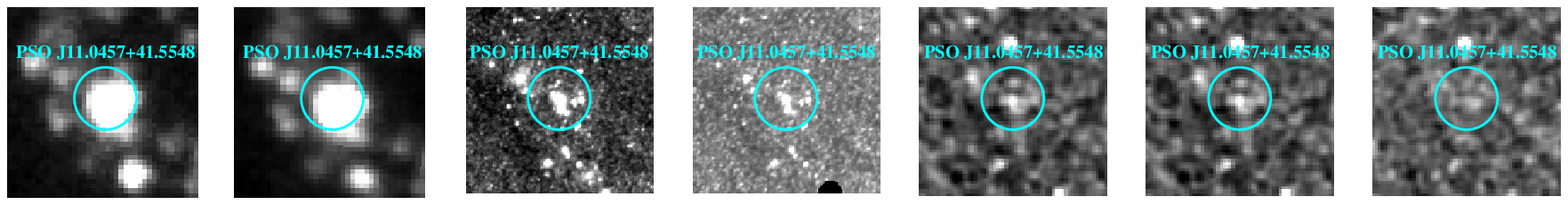}  
  \includegraphics[scale=0.9]{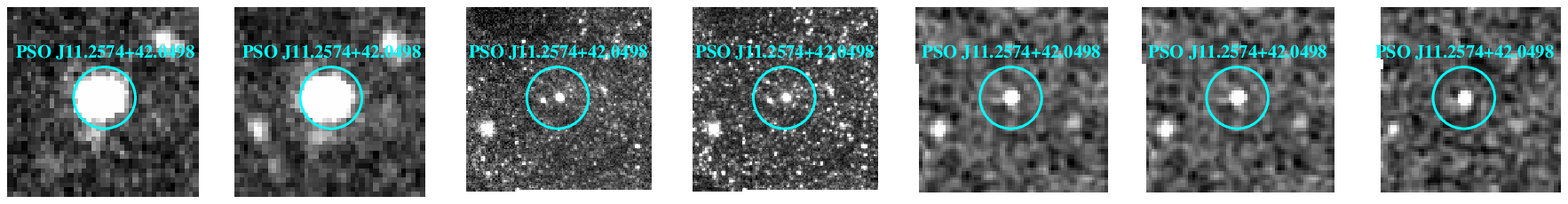}
  \caption{NUV (1516 $\AA$, first column), FUV (2267 $\AA$, second column), $\rps$ (third column), $\ips$ (forth column) and 2MASS \textit{JHK$_s$} (fifth to seventh column) postage stamps of our LBV candidates. The circles have a radius of 10''. North is to the top and east is to the left. NUV and FUV images are taken from GALEX Nearby Galaxy Atlas \citep{2007ApJS..173..185G}.}
   \label{fig.UVstamp}
\end{figure*}

\section{Discussion}

In this section we examine the properties of our 
LBV candidates, investigate whether they are UV emitters, derive their ages from the 
massive star evolutionary model, and examine their \textit{HST} images (if available).
\subsection{\textit{GALEX} UV detection}

It has been suggested that LBVs can be revealed in the UV channel \citep{1996ApJ...469..629M}. 
We have plotted the positions of our candidates on the \textit{GALEX} near-UV 
images \citep[\textit{GALEX} 
nearby galaxy atlas,][]{2007ApJS..173..185G}. As shown in Fig. \ref{fig.UVstamp}, all 
LBV candidates are aligned with bright UV sources, and they 
are located in the spiral arms of M31 (see Fig. \ref{fig.spat}). 
For comparison we also show zoom-ins of the known LBV and LBV candidates 
listed in \cite{2007AJ....134.2474M} in the appendix.

\subsection{Comparison with isochrones}

\begin{figure*}[!h]
  \centering
  \includegraphics[scale=1.0]{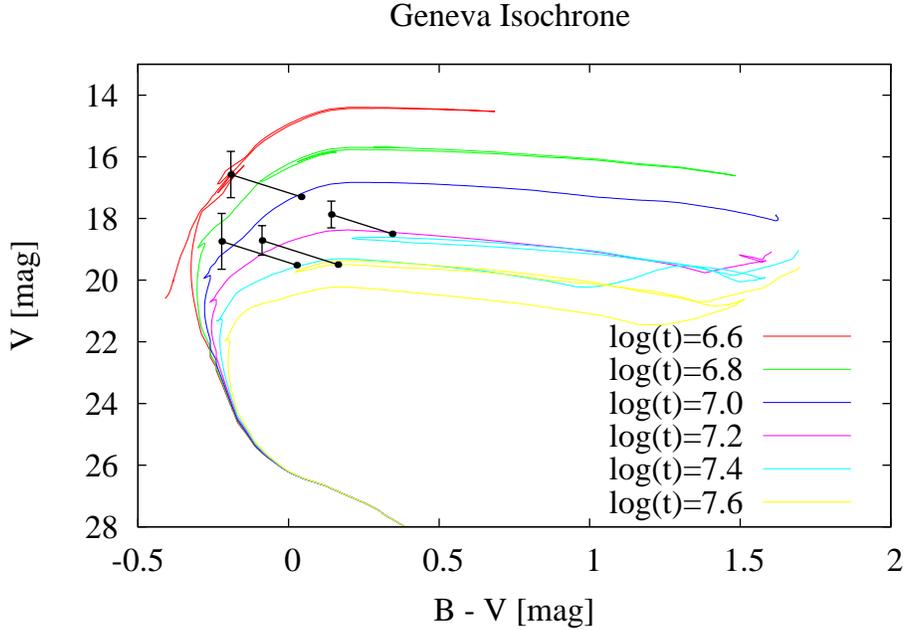}
  \caption{
Location of our LBV candidates in the B-V v.s. V CMD, over-plotted with the Geneva massive star isochrones. 
We adopt the distance modulus of 24.4 mag from \cite{1990ApJ...365..186F}. 
We apply a correction for the extinction effect on B-V color using the extinction map by \cite{2009A&A...507..283M}. 
By assuming A$_V$/E(B-V) = 3.1, we also correct the extinction effect on 
the V-band magnitude. The positions on the CMD before and after extinction correction are linked with black lines. The 
extinction corrected value are marked with error-bars.
The four 
candidates are indicated in black color. Evolutionary 
tracks of metallicity \textit{Z} = 0.014 with different age, log(t) = 6.6, 6.8, 7.0 ... 7.6 are shown with different colors 
from top to bottom. The errors on the magnitude are 
not photometric errors, but are taken from the variability seen during the time-span of PAndrome
da, as shown in Table \ref{tab.cphot}.}
   \label{fig.isochrone}
\end{figure*}

In order to see whether our candidates are consistent with the evolutionary model 
of massive stars, we compare the B-V color
and the V band magnitude of these four candidates with the latest 
Geneva evolutionary tracks \citep{2012A&A...537A.146E}.
As shown in Fig. \ref{fig.isochrone}, our candidates are in good agreement with the 
Geneva models. In addition, from the 
evolutionary tracks, we are able to estimate their ages. As indicated by the models, 
their age are of the order of 10$^7$ years.

In Fig. \ref{fig.isochrone} we also indicate the possible variabilities of LBVs by 
drawing the photometric variations $\Delta$mag$_{\rps}$ seen from 
PAndromeda as an error-bar. 
LBVs can also suffer from dust extinction from their circumstellar material.
To take this into account, we apply a correction for the extinction effect on B-V color using the extinction map by \cite{2009A&A...507..283M}. 
By assuming A$_V$/E(B-V) = 3.1, we also correct the extinction effect on 
the V-band magnitude. 
Taking the extinction effect into consideration, the true B-V value of 
our LBV candidates would be smaller. In this case, our LBV candidates 
would be blue-ward on the color-magnitude plot in Fig. \ref{fig.isochrone}, which is 
still consistent with the evolutionary model of a age in the order of 10 million years.

\subsection{HST observations}

\begin{figure}
  \centering
  \includegraphics[scale=0.7]{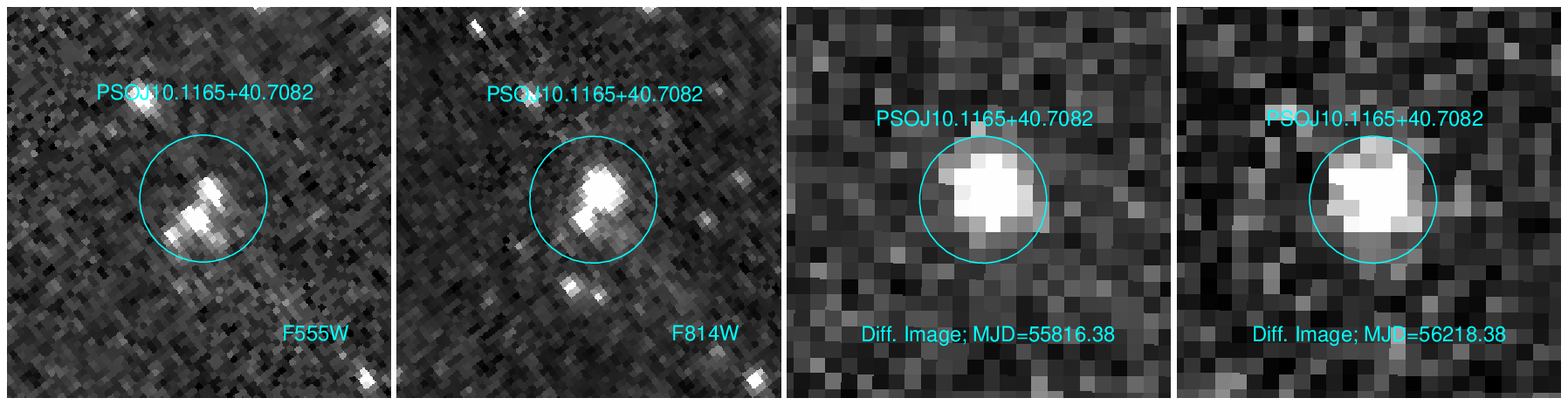}
  \includegraphics[scale=0.7]{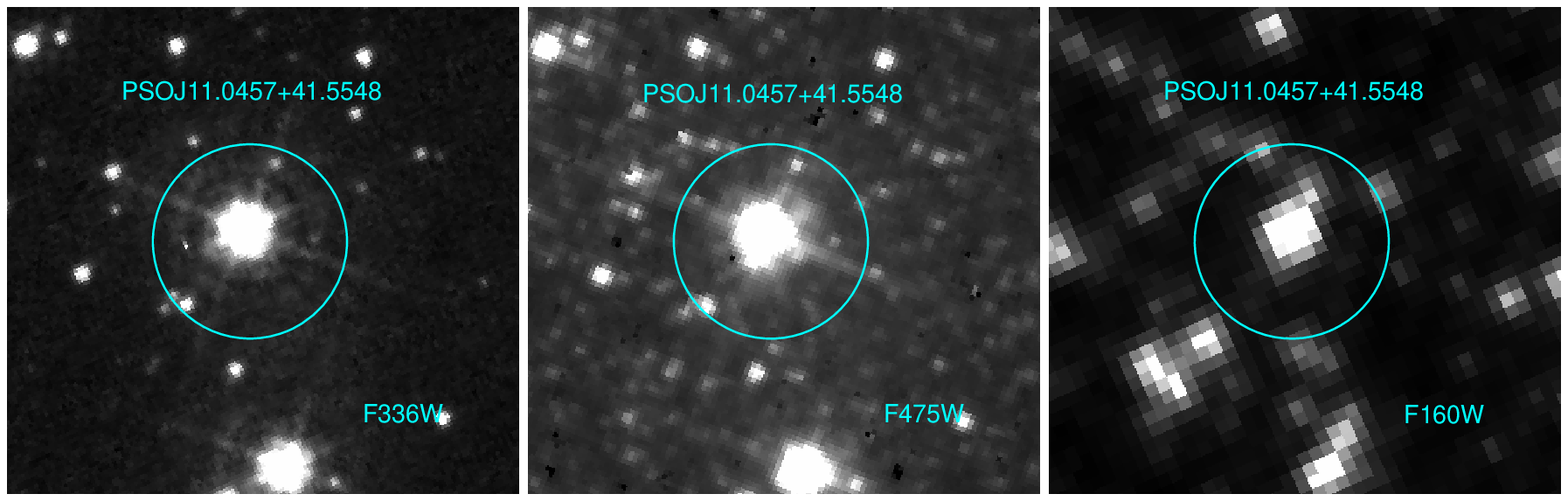}
  \includegraphics[scale=0.7]{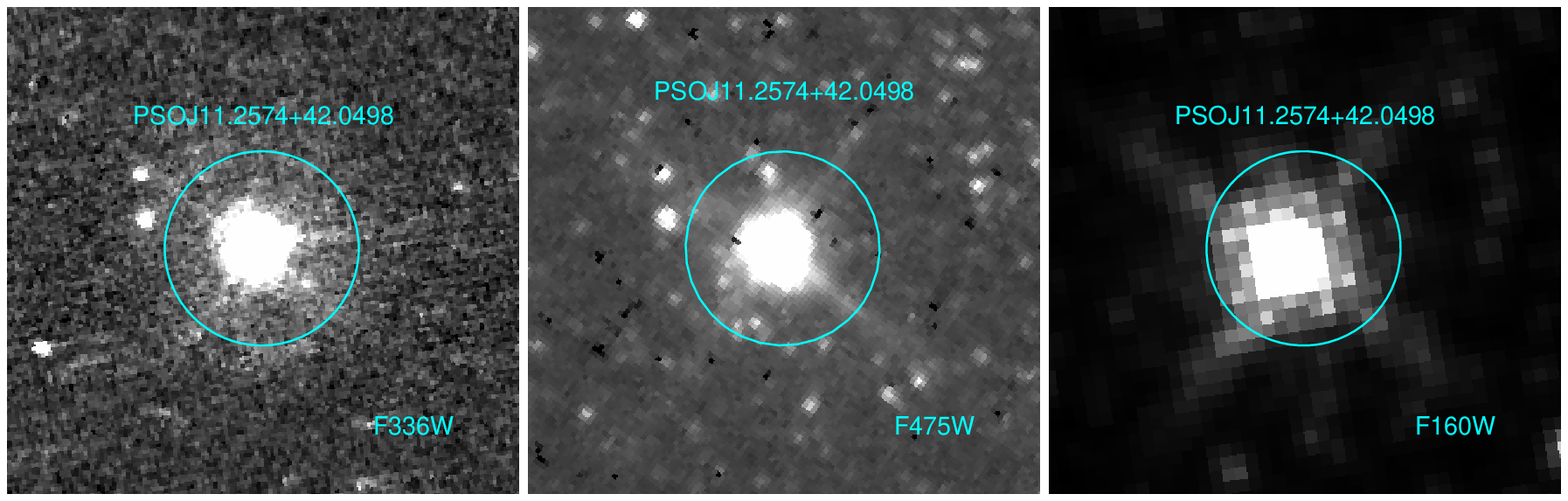}
  \caption{Postage stamps from HST archive. Upper panel: HST images of PSOJ10.1165+40.7082 from the ``Treasury Imaging of Star Forming Regions in the Local Group'' program \citep{2012AJ....144..142B}. Middle and lower panels: HST images of PSOJ11.0457+41.5548 and PSOJ11.2574+42.0498 from the ``Panchromatic Hubble Andromeda Treasury'' program \citep{2012ApJS..200...18D}. The LBV candidates are indicated by the cyan circles, 
which have a radius of 1 arcsec. The observed passbands (F160W, F336W, F475W, F555W, and F814W) are also indicated in the lower corner of each stamp. All HST images are astrometrically aligned to the PAndromeda image using our own pipeline (Kodric et al. in prep.). The median positional difference between the LGGS catalog and the PAndromeda catalog is 0.36 arcsec.}
   \label{fig.pstamphst}
\end{figure}

To confirm that our LBV candidates are stellar objects, we 
thus request the M31 HST images from the Panchromatic Hubble Andromeda Treasury project \citep[PHAT,][]{2012ApJS..200...18D}. 
Since the PHAT survey only covers the northern disk of M31, 
we only find images
for two of our LBV candidates, PSO J11.0457+41.5548 and PSOJ11.2574+42.0498. We show the HST ACS images of them in Fig. \ref{fig.pstamphst}.
With the exquisite spatial resolution of HST, we can see that the shape of PSO J11.0457+41.5548 and PSOJ11.2574+42.0498 traces the typical point-spread function in the field.

In addition to the PHAT archived images, we also found PSO J10.1165+40.7082 covered by the ``Treasury Imaging of Star Forming Regions in the Local Group'' program \citep{2012AJ....144..142B}. The HST images of this candidate, astrometrically aligned to the PAndromeda data using our own pipeline (Kodric et al. in prep.), is shown in Fig. \ref{fig.pstamphst} as well. HST resolved two sources within 1 arcsec of PSO J10.1165+40.7082. To distinguish which is the varying source, we examine the difference image from PAndromeda during maximum flux (at MJD=55816.38 and 56218.38) and found that the brighter source in the F814W band is the varying source.

\subsection{A further look at the variability criterion}
\label{sec.det_eff}
\begin{figure*}[!h]
  \centering
  \includegraphics[scale=1.0]{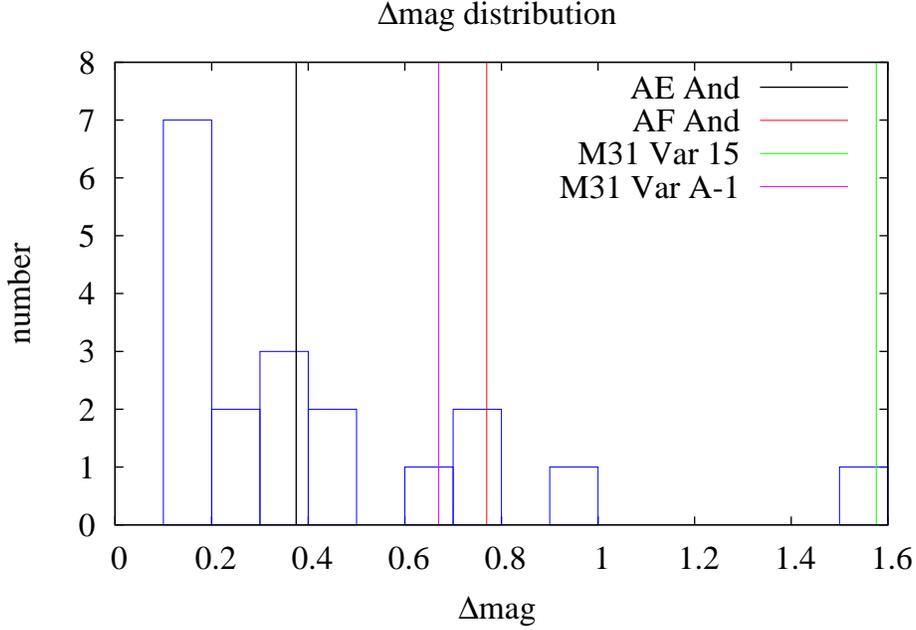}
  \caption{$\Delta$mag$_{\rps}$ distribution of variable sources passing our optical and infrared color criteria: 
we show the number of lightcurves which fulfill 
both the H-K$<$0.5,V$<$20, and B-V $<$ 0.5 criteria. Seven lightcurves pass these criteria, and four of them are previously 
unknown. The rest of them are known LBVs.
The vertical lines indicate the variability of known LBVs from \cite{2006AJ....131.2478M}.}
   \label{fig.det_eff}
\end{figure*}

In  Fig. \ref{fig.det_eff}, we plot the number of sources which pass our optical 
and infrared photometric criteria against the photometric variability from PS1 
$\rps$-band lightcurves. There are in total seven 
sources showing $\Delta$mag$_{\rps}$ $>$ 0.4 mag out of which we 
selected four as possible LBV candidates.
Among the remaining three, they are all known 
LBVs (AF And, M31 Var 15, and M31 Var A-1).
In Fig. \ref{fig.det_eff}, there are three more sources that varying at 0.3 mag level, which
are AE And with $\Delta$mag$_{\rps}$ = 0.375 and other variables with $\Delta$mag$_{\rps}$ = 0.339 and 0.307. Even 
if we lowered the $\Delta$mag$_{\rps}$ criterion to the lowerest value of known LBVs (0.375), we would
only select AE And, but no additional new LBV candidates.

For comparison, we also show the $\rps$ lightcurves of the four known LBVs listed in \cite{2007AJ....134.2474M} in the appendix.

\section{Summary and Outlook}
We study the photometry of known M31 LBVs from \cite{2007AJ....134.2474M} and present a new 
approach to search for LBVs using optical and infrared information.
We have selected four candidate LBVs sharing the same properties of 
known LBVs in terms of optical and infrared colors; 
they are 
also observed in the \textit{GALEX} UV data and all of them are 
located within M31 spiral 
arms. Our sample exhibits photometric variation $>$ 0.4 mag as 
seen from the PAndromeda 
survey. These sources are in agreement with the stellar evolution model of the 
Geneva group, which 
gives an age between 10 and 100 million years. This implies that while low mass 
stars are still in the early evolution stage in the spiral arms, massive stars 
have already evolved into LBV stage.
Though the true nature of our sample awaits a spectroscopic confirmation, the 
bright UV emission, optical and infrared colors, photometric variability, 
and the HST image altogether indicate
that our candidates are very likely a LBV. 

We will request spectroscopic observation to confirm our candidates as true LBVs, and 
classify them according to the taxonomy scheme outlined by \cite{2007AJ....134.2474M}. 
In addition, it has been shown that LBVs can be surrounded by 
nebula with dust \citep{2013A&A...557A..20V}.
Spectra in the mid infrared will help us to 
probe dusts with colder temperature; 
\cite{1997ASPC..120..326W} have used the Short Wavelength Spectrometer on-board ISO 
to obtain mid-infrared spectra of several known LBVs and led to the 
discovery of cold circumstellar dust with a temperature about $\sim$ 50K. 
Future space observatories like \textit{SPICA} telescope \citep{2001ESASP.460..475N} will 
also provide spectra in the wavelength of mid-infrared.

\section{Acknowledgments}
We would like to thank the referee for useful comments. This work was supported by the DFG cluster of excellence ‘Origin and Structure of the
Universe’ (www.universe-cluster.de).

The Pan-STARRS1 Surveys (PS1) have been made possible through contributions of the Institute for Astronomy, the University of Hawaii, the Pan-STARRS Project Office, the Max-Planck Society and its participating institutes, the Max Planck Institute for Astronomy, Heidelberg and the Max Planck Institute for Extraterrestrial Physics, Garching, The Johns Hopkins University, Durham University, the University of Edinburgh, Queen's University Belfast, the Harvard-Smithsonian Center for Astrophysics, the Las Cumbres Observatory Global Telescope Network Incorporated, the National Central University of Taiwan, the Space Telescope Science Institute, the National Aeronautics and Space Administration under Grant No. NNX08AR22G issued through the Planetary Science Division of the NASA Science Mission Directorate,  the National Science Foundation under Grant No. AST-1238877, and the University of Maryland.

This publication makes use of data products from the Two Micron All Sky Survey, which is a joint project of the University of Massachusetts and the Infrared Processing and Analysis Center/California Institute of Technology, funded by the National Aeronautics and Space Administration and the National Science Foundation.

This research has made use of the NASA/ IPAC Infrared Science Archive, which is operated by the 
Jet Propulsion Laboratory, California Institute of Technology, under contract with the 
National Aeronautics and Space Administration.

\textit{GALEX} (Galaxy Evolution Explorer) is a NASA Small Explorer, launched in April 2003. We 
gratefully acknowledge NASA's support for construction, operation, and science analysis for the GALEX 
mission, developed in cooperation with the Centre National d'Etudes Spatiales of France and the Korean 
Ministry of Science and Technology.

This research has made use of the SIMBAD database,
operated at CDS, Strasbourg, France

\clearpage
\section{Appendix}
In this section, we present information of LBV and LBV candidates from \cite{2006AJ....131.2478M}, as supporting materials 
to show that our LBV candidates bear similar properties to known LBVs.

In Table \ref{tab.knownLBV}, we collect the V-band magnitude, the B-V color, the photometric variability in the $\rps$ lightcurve
of four known LBVs in the \cite{2007AJ....134.2474M} sample.
\begin{table*}
\center
\begin{tabular}[t]{lllll}
\hline 
Name   & LGGS &       V      &  B-V    & $\Delta$m in $\rps$  \\
       & nomenclature    &              &         & [mag]     \\
\hline
AE And      & J004302.52+414912.4 & 17.426$\pm$0.005 & -0.153$\pm$0.005  & 0.375055  \\        
AF And      & J004333.09+411210.4 & 17.325$\pm$0.004 &  0.013$\pm$0.004  & 0.769598  \\
M31 Var 15  & J004419.43+412247.0 & 18.450$\pm$0.004 & -0.007$\pm$0.004  & 1.575990  \\         
M31 Var A-1 & J004450.54+413037.7 & 17.143$\pm$0.004 &  0.211$\pm$0.005  & 0.669998  \\        
\hline
\end{tabular}
\caption{Properties of known LBVs.} 
\label{tab.knownLBV}
\end{table*}

\subsection{GALEX UV detection}
\begin{figure}
  \centering
  \includegraphics[scale=0.7]{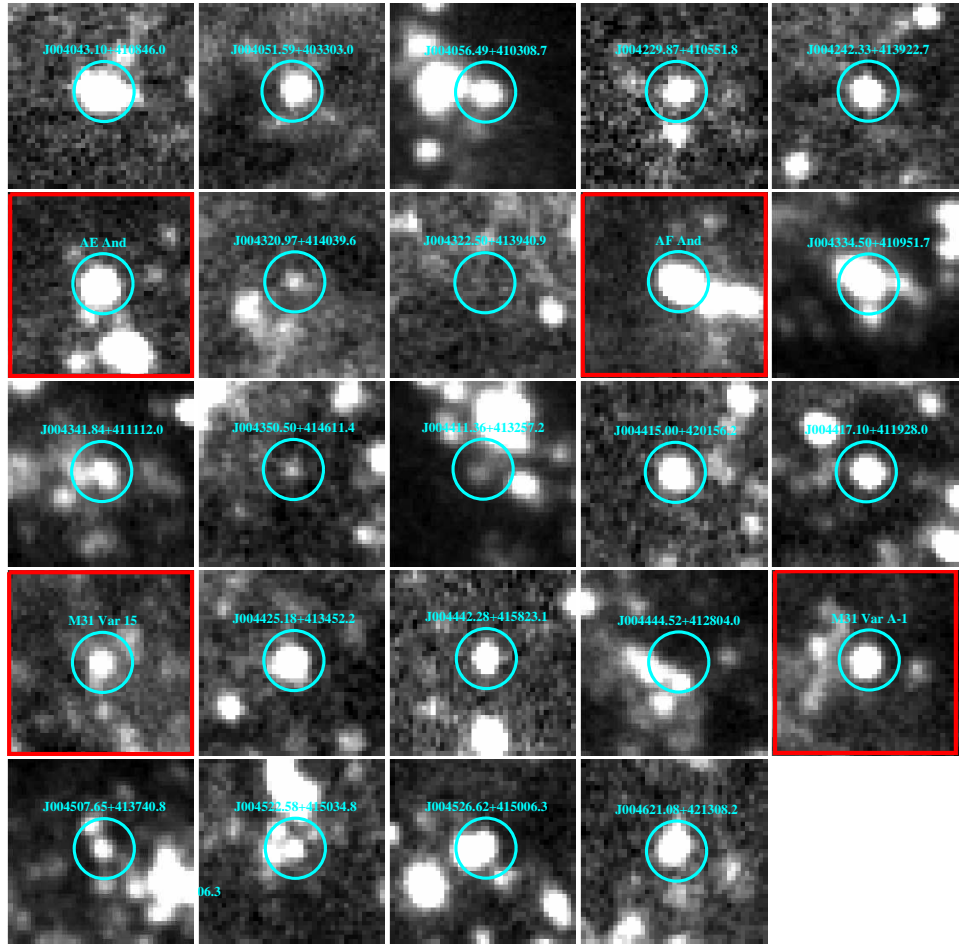}
  \caption{NUV postage stamps of the LBV and candidate LBV sample from \cite{2007AJ....134.2474M}. The red squares outline the four known M31 LBVs. The images are taken from the \textit{GALEX} Nearby Galaxy Atlas.}
   \label{fig.nuvpstampM07}
\end{figure}

\begin{figure}
  \centering
  \includegraphics[scale=0.7]{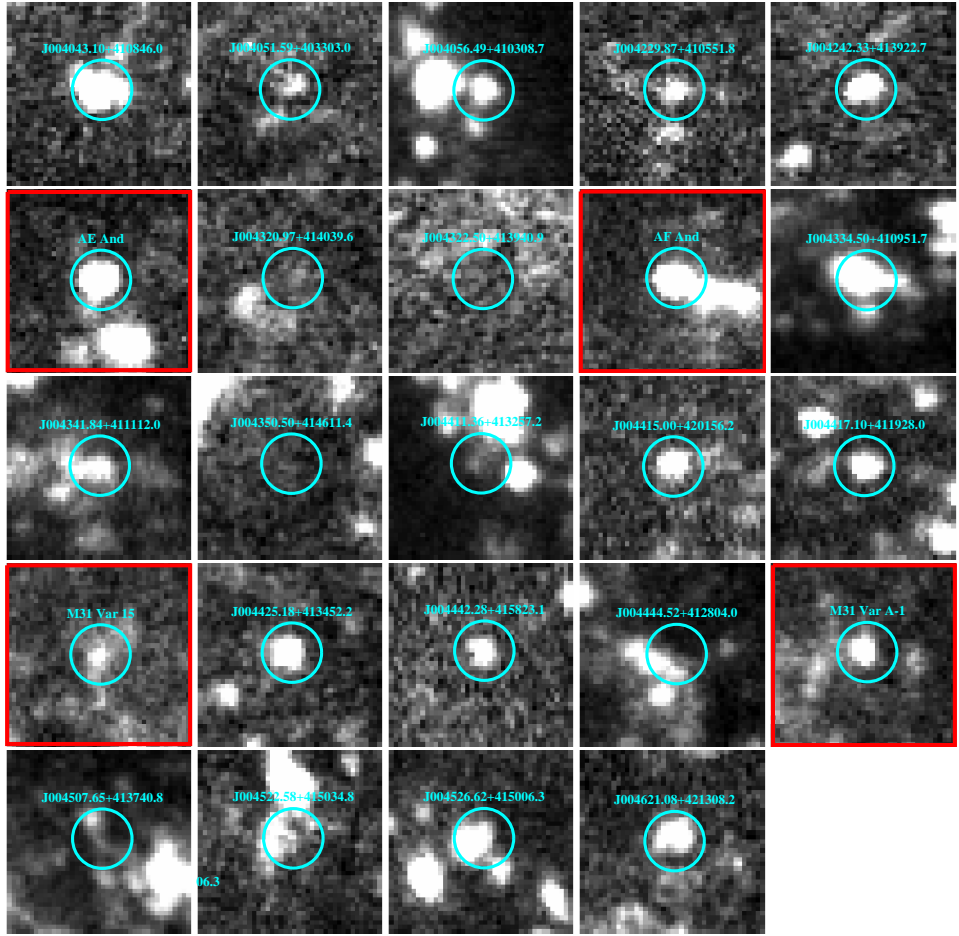}
  \caption{FUV postage stamps of the LBV and candidate LBV sample from \cite{2007AJ....134.2474M}. 
The red squares outline 
the four known M31 LBVs. The images are taken from the \textit{GALEX} Nearby Galaxy Atlas.}
   \label{fig.fuvpstampM07}
\end{figure}

We collect UV images of known LBV and LBV candidates listed in \cite{2007AJ....134.2474M} and present in Fig. \ref{fig.nuvpstampM07} and \ref{fig.fuvpstampM07}.
The images are taken at near and far UV by \textit{GALEX}. Most of the LBV and LBV candidates in \cite{2007AJ....134.2474M} are 
bright sources in NUV, but some of them show faint or no FUV emission.

\subsection{HST observations}

\begin{figure}
  \centering
  \includegraphics[scale=0.7]{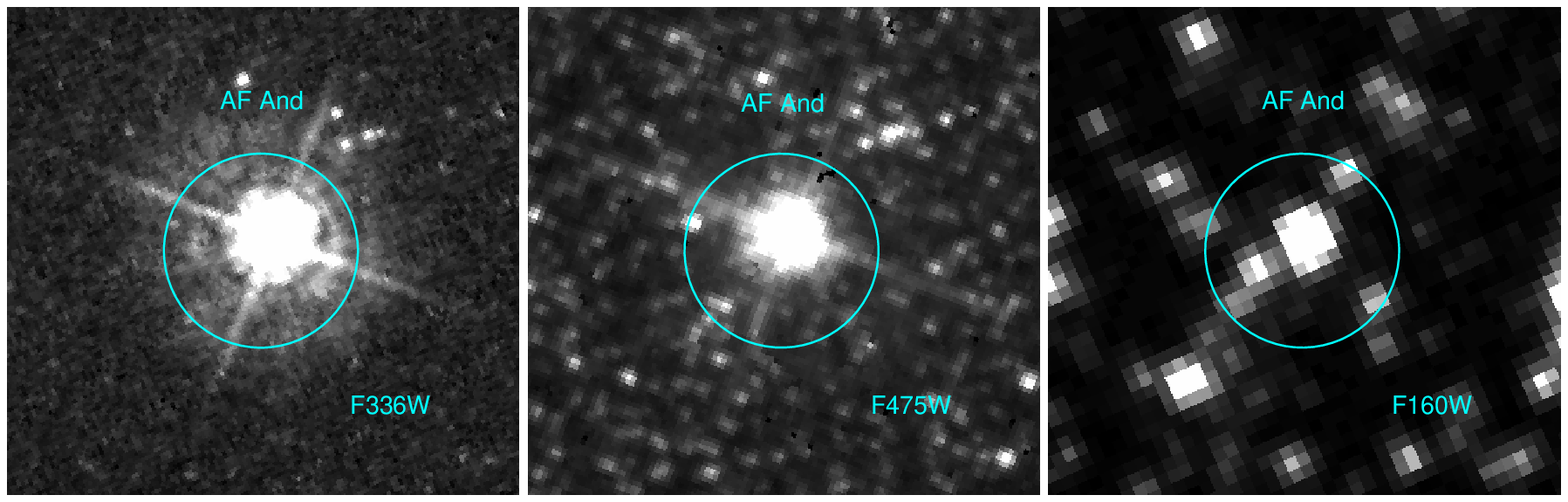}
  \includegraphics[scale=0.7]{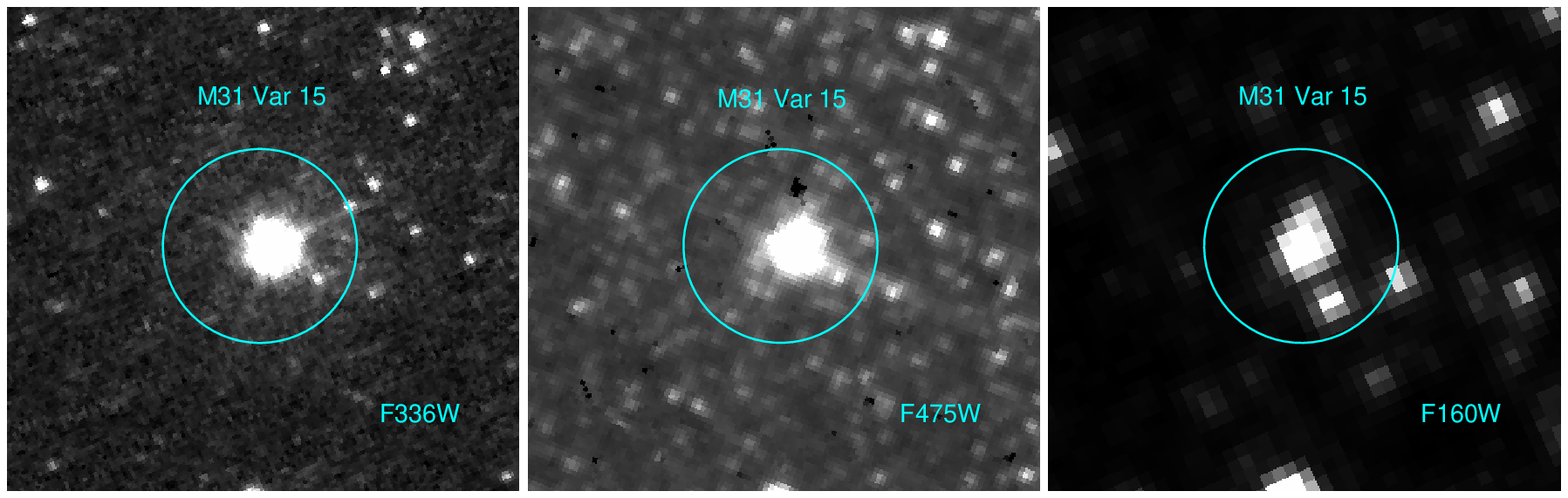}
  \caption{Postage stamps of the HST observations from \cite{2012ApJS..200...18D}. The upper and lower panel
show the HST images of AF And and M31 Var 15, respectively. The LBVs are indicated by the cyan circles, 
which have a radius of 1 arcsec. The observed passbands (F160W, F336W, and F475W) are also indicated in the lower corner of each stamp.}
   \label{fig.pstamphst3}
\end{figure}

We search for HST archive images of the known LBVs listed in \cite{2007AJ....134.2474M} in the 
``Panchromatic Hubble Andromeda Treasury'' program \citep{2012ApJS..200...18D}, and found
two of the known LBVs (AF And and M31 Var 15) have been observed by this program.
Their images are shown in Fig. \ref{fig.pstamphst3}.

\subsection{Lightcurves from PS1 data}
\begin{figure*}
  \centering
  \includegraphics[scale=0.52]{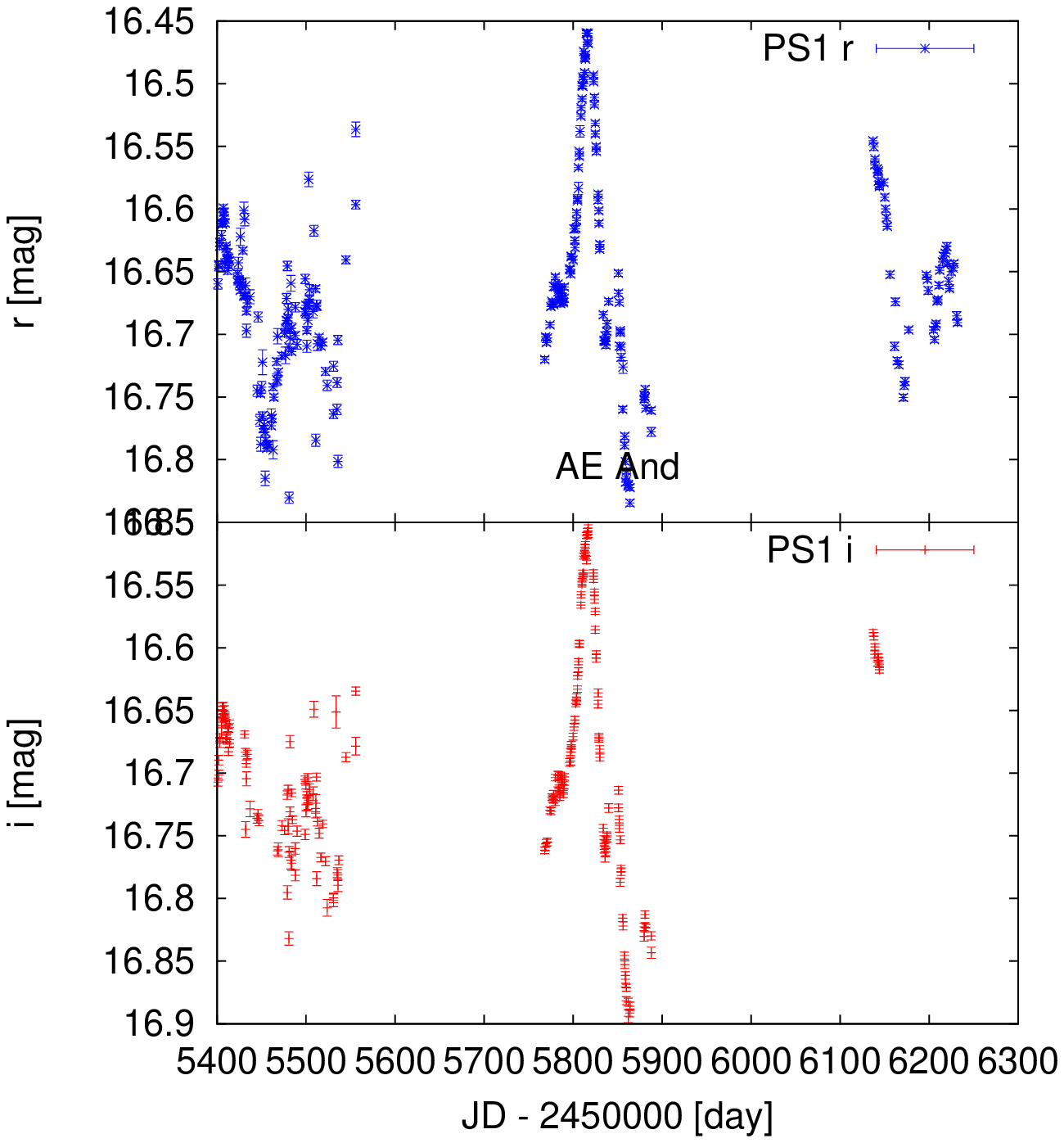}
  \includegraphics[scale=0.52]{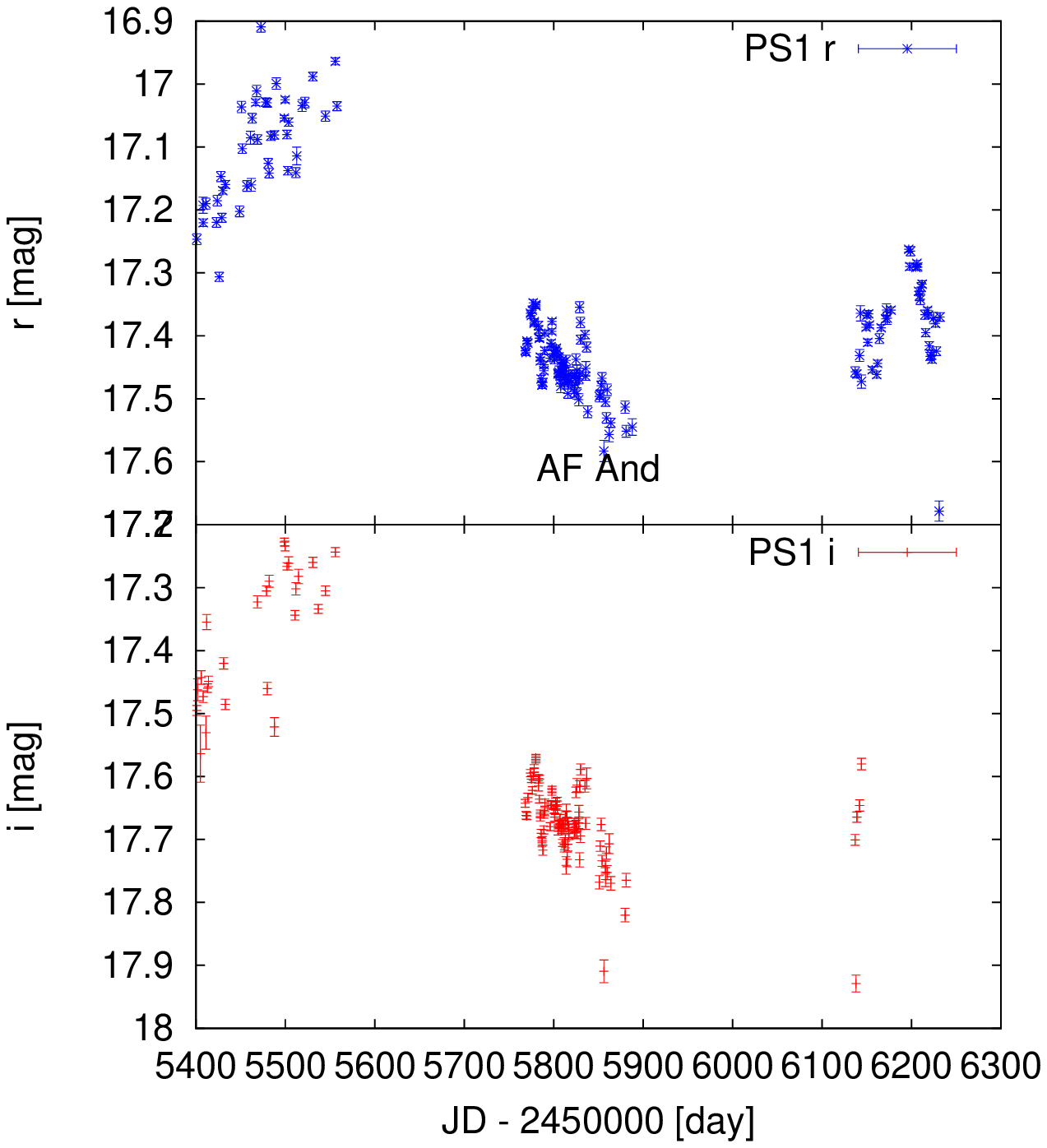}
  \includegraphics[scale=0.52]{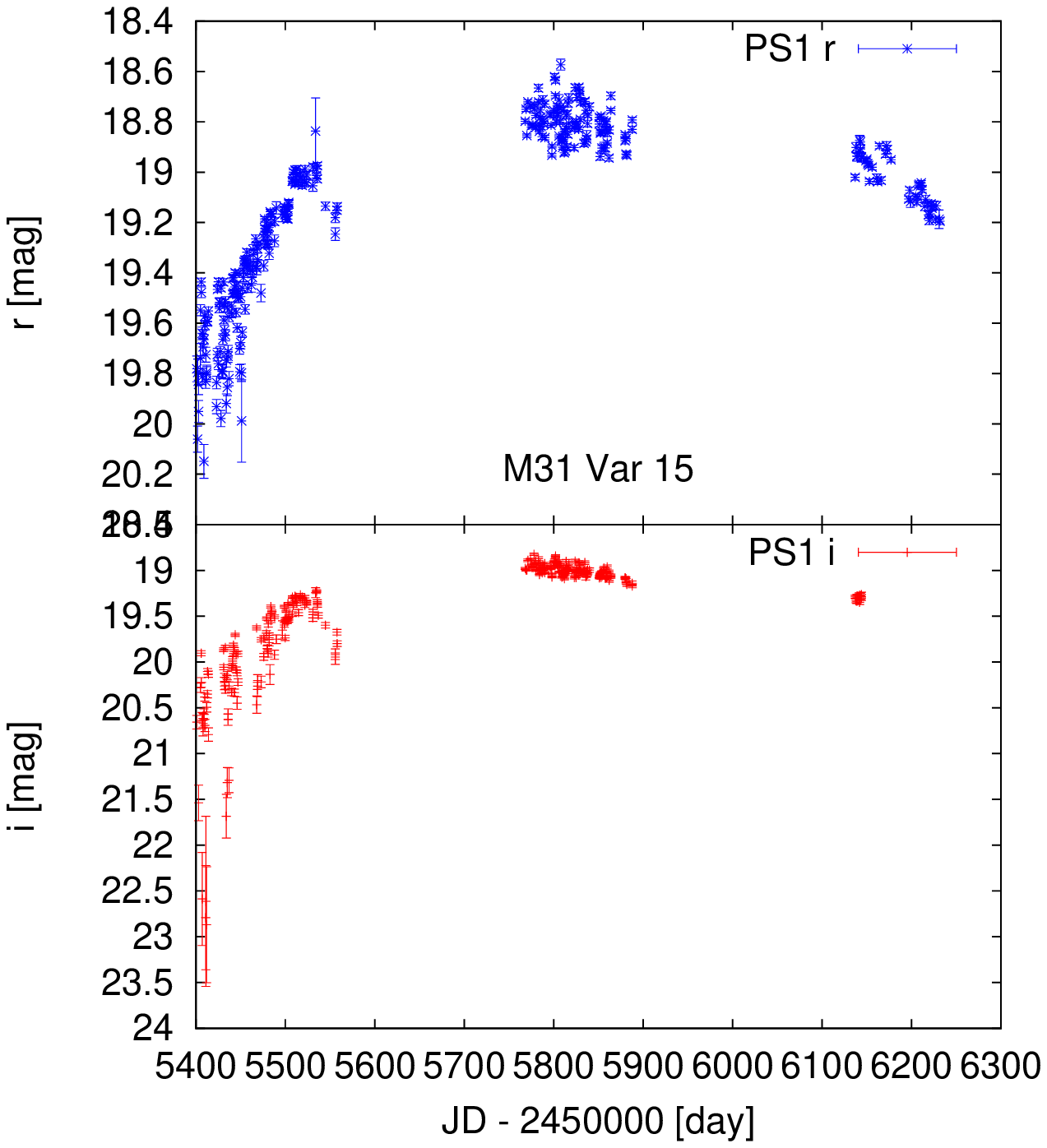}
  \includegraphics[scale=0.52]{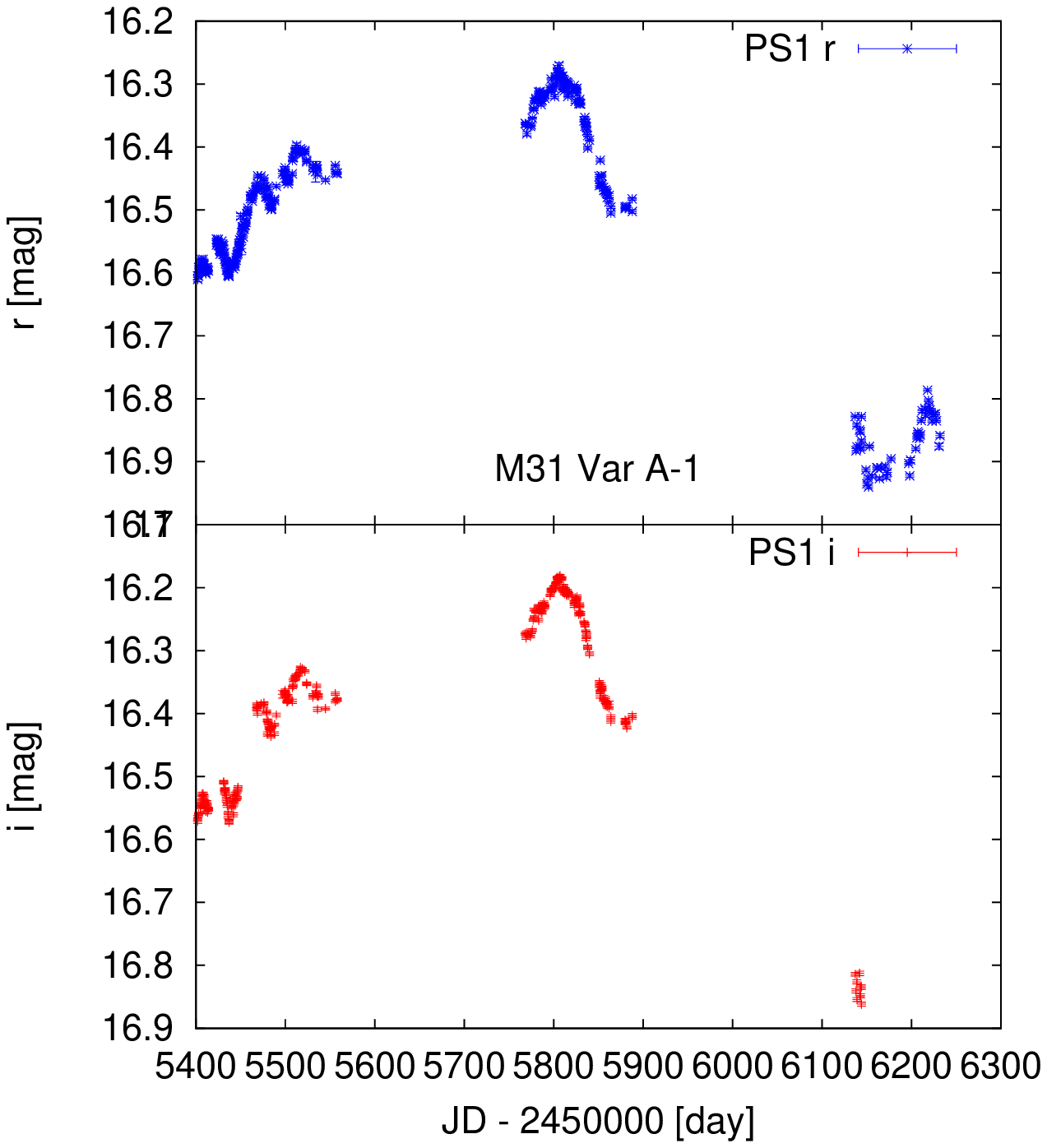}
  \caption{PS1 lightcurves of known LBVs from \cite{2007AJ....134.2474M}.}
   \label{fig.opt1}
\end{figure*}

We show the lightcurves of four known LBVs from \cite{2007AJ....134.2474M} in Fig. \ref{fig.opt1}. The lightcurves are from
our PS1 three year data. The photometric variation during the three year time-span are listed in Table \ref{tab.knownLBV}.

\end{document}